# SERVICE COMPOSITION DESIGN PATTERN FOR AUTONOMIC COMPUTING SYSTEMS USING ASSOCIATION RULE BASED LEARNING AND SERVICE-ORIENTED ARCHITECTURE

Vishnuvardhan Mannava[1] and T. Ramesh[2]

[1]Department of Computer Science and Engineering,
K L University, Vaddeswaram, 522502, India
`vishnu@kluniversity.in`
[2]Department of Computer Science and Engineering,
National Institute of Technology, Warangal, 506004, India
`rmesht@nitw.ac.in`

## ABSTRACT

*In this paper we will compose the design patterns which will satisfy the properties of autonomic computing system: for the Decision-Making phase we will introduce Case-Based Reasoning design pattern, and for Reconfiguration phase we will introduce Reactor design pattern. The most important proposal in our composite design pattern is that we will use the Association Rule Learning method of Data Mining to learn about new services that can be added along with the requested service to make the service as a dynamic composition of two or more services. Then we will include the new service as an aspectual feature module code without interrupting the user. As far as we know, there are no studies on composition of design patterns and pattern languages for autonomic computing domain. We will authenticate our work by a simple case study work. A simple Class and Sequence diagrams are depicted.*

## KEYWORDS

*Autonomic System, Design Patterns, Aspect-Oriented Design Patterns, Feature-Oriented Programming (FOP), Aspect-Oriented Programming (AOP), Data Mining*

## 1. INTRODUCTION

Pattern composition has been shown as a challenge to applying design patterns in real software systems. composite patterns represent micro architectures that when glued together could create an entire software architecture. Thus pattern composition can lead to ready-made architectures from which only instantiation would be required to build robust implementations. A composite design patterns shows a synergy that makes the composition more than just the sum of its parts. As far as we know, there are no studies on composition of design patterns and pattern languages for autonomic computing domain. The most widely focused elements of the autonomic computing systems are self-* properties. So for a system to be self-manageable they should be self-configuring, self-healing, self-optimizing, self-protecting and they have to exhibit self-awareness, self-situation and self monitoring properties [4]. For providing dynamic behavior in currently developed systems some of the newly introduced programming language features are required. The most popular and interesting re-search area in providing dynamic adaptability in today's programming world is with Aspect-Oriented Programming (AOP) [16] [2] and Feature-Oriented Programming (FOP) [5]. Design patterns are most often used in developing the software system to implement variable and reusable software with object oriented programming (OOP) [16].Most of the design patterns have been successfully developed in the OOPs, but at the same time developers have faced some problems like as said in [12] they found the lack of modularity, composability and reusability in respective object oriented designs [9]. The cause of problem in Applying OOPs in developing design patterns was found due to the crosscutting concerns. Crosscut-ting concerns are the problems that result in code tangling, scattering, and replication of code when software is decomposed along one dimension [19]. So in order to overcome this problem some advanced modularizing techniques have been introduced like AOP and FOP. With the help of Aspect-Oriented Programming we can separate the crosscutting concerns from the main functional logic. We can handle these crosscutting concerns in separate Aspect and pointcut like modules. On the other hand the Feature-Oriented Programming has made the way for research in software-product lines with the inclusion of new features in already developed software as a separate Feature Module by making the life of developers easier by not develop again from scratch. In this paper we will propose a design pattern that will develop an autonomic system that will use Case-Based Reasoning Design Pattern [15] for Decision-Making, and for Reconfiguration phase we will use Reactor Design Pattern [6]. With the help of the Aspectual Feature Module based technique proposed in [2] we can include the new service into the existing system as a new Feature. Here we have concentrated on the Learning process with the help of Data Mining methods. The Association Rule Learning in data mining is the main concept that we have used to create a Dynamic composition services from the requested service. To understand this let's look at an example, if a customer visits a site and requests for a service say he buys a computer and then after that he will also buy the UPS like this different customers who buy the computer will also buy an UPS and also printer. So with our pattern these transactions are stored in the Learning data base which will be used by the Association rule method of data mining to determine a rule like computer, UPS =¿ printer means the customer who buys a computer and at the same time UPS, he will also buy a printer, with the help of Association rule data mining instead of requesting the three items individually we can dynamically compose service which is a composition of three services as a single service. So when a customer who comes to buy the three items at a single request can just access this composed service and no need to request three items separately. Then we will just include this new composed service a Feature into the Service Repository so that the plan to take decisions regarding like this services will be easy. By providing the Learning process with association learning rule of data mining we can dynamically reconfigure the system to be adaptive to the new services. Refer Figure 1 for the Autonomic computing system in proposed in [15].

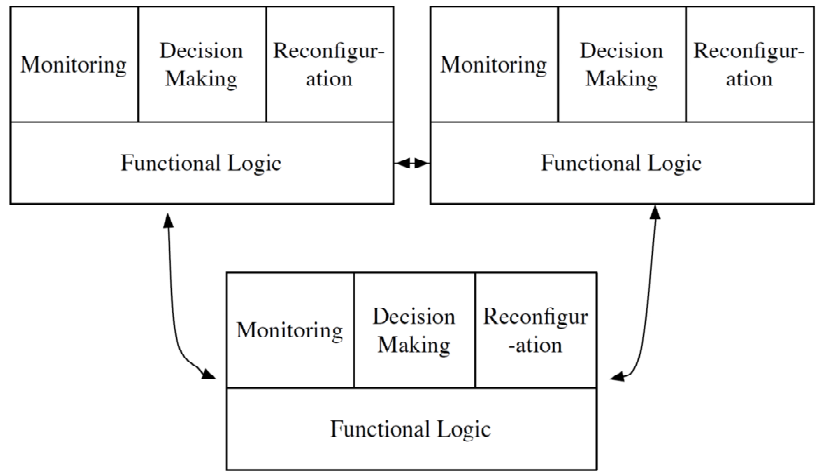

Figure 1: Autonomic Computing System

## 2. RELATED WORK

In this section we present some works that deal with different Design patterns oriented work. There are number of publications reporting the reusability features of the design patterns when the same problem occurs in the future.

Demian Antony D'Mellon, V.S. Ananthanarayana, and Supriya Salian in paper [7] proposed the review of Web Services composition Architectures and techniques used to generate new services.

V.S.Prasad Vasireddy, Vishnuvardhan Mannava, and T. Ramesh paper [14] discuss applying an Autonomic Design Pattern which is an amalgamation of chain of responsibility and visitor patterns that can be used to analyze or design self-adaptive systems. They harvested this pattern and applied it on unstructured peer to peer networks and Web services environments.

Olivier Aubert, Antoine Beugnard [13] they proposed an Adaptive Strategy Design Pattern that can be used to analyze or design self-adaptive systems. It makes the significant components usually involved in a self-adaptive system explicit, and studies their interactions. They show how the components participate in the adaptation process, and characterize some of their properties.

In Vishnuvardhan Mannava, and T. Ramesh paper [21] they have proposed a design pattern for Autonomic Computing System which is designed with Aspect-oriented design patterns and they have also focused on the amalgamation of the Feature-oriented and Aspect-oriented software development methodology and its usage in developing a self-reconfigurable adaptive system.

In Vishnuvardhan Mannava, and T. Ramesh paper [22] they have proposed a system for dynamically configuring communication services. Server will invoke and manage services based on time stamp of service. The system will reduce work load of sever all services in executed by different threads based on time services are executed, suspended and resumed.

In Vishnuvardhan Mannava, and T. Ramesh paper [23] they have proposed an adaptive reconfiguration compliance pattern for autonomic computing systems that can propose the reconfiguration rules and can learn new rules at runtime.

In Sven Apel, Thomas Leich, and Gunter Saake [2] they proposed the symbiosis of FOP and AOP and aspectual feature modules (AFMs), a programming technique that integrates feature modules and aspects. They provide a set of tools that support implementing AFMs on top of Java and C++.

The authors in paper [21] have mainly concentrated on providing the adaptability to the application at runtime by using Aspect-Oriented Programming. Here we are concerned with the Feature Oriented Programming Module used by them, that is the authors have developed there system/application with the help of FeatureIDE for providing the amalgamation of AOP and FOP in the application/System. We would like to introduce the same Feature-Oriented modularization with the help of a pure an extensible compiler for Feature-Oriented Programming in java called "Fuji". With this we want to convey that we would like to enhance the adaptable nature of the design pattern proposed in paper [21] by applying Fuji compiler to provide more efficient Feature-Oriented software development.

At the same time the authors in the paper [14] have mainly concentrated on the development of a peer-to-peer unstructured distributed application. They have only discussed about the invocation of the web services in a peer-to-peer distributed environment. The papers [14] [13] gave use the inspiration of how the design patterns amalgamation can lead to a more complex design patterns that may solve the problems in a server or distributed environment. We would like to enhance the design pattern introduced by the authors in [14] by applying pure socket programming and Java Web Services (JWS) for the purpose of web services invocation and also we use the Association Rule Learning Appriori algorithm for developing a self-learning system which is capable of developing new service which are the compositions of already existing services.

In the current literature there are only applications/systems which can either solve only one task that is either self-configuration in distributed environment or self-configuration in central with in server environment, but there is a gap between the development of a composed model of these two modules. So we are introducing this proposal to fill that gap and to develop the application which have the capability to take decisions at run time to decide wheather the computation is to be done as a distributed computation with SOA or with in server computation by using aspect-oriented programming or feature-oriented programming.

## 3. PROPOSED AUTONOMIC DESIGN PATTERN

One of the objectives of this paper is to apply the Association rule based Learning method of Data Mining to determine the new service and include it as a Aspectual Feature-oriented code into the existing system. Initially the customer will request for a service then the Trigger class will trigger an event to the server. With the help of the Case-Based Reasoning Design Pattern [15] we can take a decision that will decide which plan should be applied to provide the service to the user. Here the Decision is taken based on the predefined rules in the Fixed Rules class then these rules are given as input to the Decision class by the Inference class. After a service is planned the data related to this transaction will be stored in the trigger repository for future use and also in the Learning Repository. Once the Service providing plan is selected then it will be given as input to the Reactor Design Pattern [6] which intern provides the service to the customer with a Service Handling Mechanism in Reactor Pattern. All the services provided or handled in the reactor pattern are provided in a distributed manner. A service is provided by other servers in the distributed environment by establishing the connections with the peer servers using the Service-Oriented Architecture (SOA).

On the other hand by performing the Association Rule based learning of data mining on the data stored in Learning Repository the server will generate new service which is composition of two or more services. So the Association Rule learning helps to include a new service with the help of Aspectual Feature module code in to the already existing Service Repository. So that more efficient access of services can be provided to the customers and all this process is done at run time without the inclusion of a developer manually, means self reconfiguration is provided at run-time. So one thing to focus here is that the service which is composition of two or more can provide the service logic present with in the current central server environment or it can provide the service at a remote server. In this way the server can use the services provided within the server without the need to contact the peer server for providing service and reducing the overhead of distributed computing or if the service is not available with the current server then it may switch to distributed computing mode.

A design pattern is a particular form of recording information about a design such that the same pattern can be applied in future, if same situation is repeated to solve problem. So the design patterns are accepted in wide range of object-oriented designs. Collections of design patterns can be found in numerous publications. Some of the design patterns that are employed to in the autonomic system are described below.

# 4. DESIGN PATTERN TEMPLATE

To facilitate the organization, understanding, and application of the proposed design patterns, this paper uses a template similar in style to that used in [15].

## 4.1 Pattern Name

Service Composition Design Pattern.

## 4.2 Classification

Decision-Making.

## 4.3 Intent

Systematically applies the Design Patterns to an Autonomic Computing System and insertion of a new dynamically composed service interns of a Aspectual Feature Module (AFM) [2] with the help of Association rule based Learning method of data mining.

## 4.4 Context

Our design pattern may be used when:

- Self-learning type of autonomic computing property has to be achieved.

- For dynamically composing web services and to generate a new service which will perform the complex tasks of the customers.
- To include the newly discovered services as a Feature Module into the currently running system and satisfy the self-reconfiguration property.

## 4.5 Proposed Pattern Structure

A UML class diagram for the proposed design Pattern can be found in Figure 2.

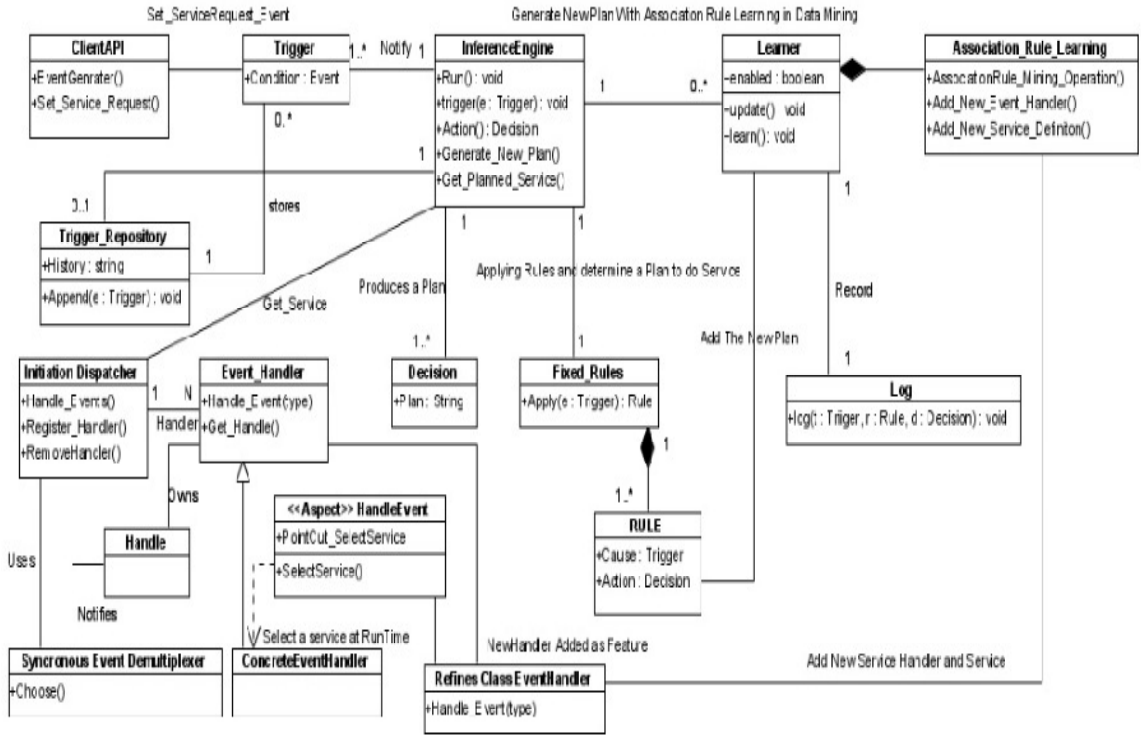

Figure 2: Class Diagram for the proposed composite Pattern

## 4.6  Participants

- ClientAPI: The client will supply the service he wants to access and the CleintAPI class will generate a Event to the Trigger class to serve the Requested service.

- Trigger: The trigger class will notify the Inference Engine that an event has been generated and it has to handle a requested service.

- Inference Engine: This class will just takes the event generated by trigger class and it will check whether a rule that can select a correct plan exists in the Decision class.

- Fixed Rules: This class will check for any rules that satisfy the given service request and then it will respond with a rule that satisfies most.

- Decision: After a rule is selected by the Inference Engine it will select the Correct Plan that can fulfill the service requested by the customer with Decision class.

- TriggerRepositroy: Here this class will store the details of the event generated and the cause of the event and also timestamp like that it store some information.
- Learner: This class will store all the service transactions that have taken place in the server and it will be used for future purpose to derive new services and for service composition. When new service is composed in the Association Rule Learning it will also add the new rule of that respective Service.

- Association Rule Learning: This is the important class where it will perform the Association Rule based learning method of the Data Mining. It will access the information stored in the Learner class and perform the Association rule based learning on that information and then it will generate the new services which are composited of two or more service.

- Log: it will store the trigger, Rule, Decision Related information for the record storing purpose.

- Initiation Dispatcher: It defines an interface for registering, removing, and dispatching the event handlers.

- Synchronous Event Demultiplexer: The synchronous Event Demultiplexer is responsible for waiting until new events occur. When it detects new events, it will inform the Initiation Dispatcher to call back application-specific event handler.

- Event Handler: Specifies an interface consisting of a hook method [6] that abstractly represents the dispatching operation for service-specific events. This method must be implemented by application-specific services.

- Concrete Event Handler: Implements the hook method [6], as well as the methods to process these events in an application-specific manner. Applications register Concrete Event Handlers with the Initiation Dispatcher to process certain types of events. When these events arrive, the Initiation Dispatcher calls back the hook method of the appropriate Concrete Event Handler.

- Handle Event Aspect: This is the aspect-oriented implementation module that will be viewed into the concrete event handling class, so that only a particular requested service method get viewed into the code at runtime.

- Refines Class Event Handler: This will add a new event handler for a new service and also the composed service in such a way that the insertion will be done as a new feature with the help of FOP.

## 4.7 Consequences

- This design pattern will eliminate the number of service requests that should be sent to the server.

- With the help of Association Rule based Learning we can easily achieve the Dynamic Service Composition Techniques.

- We can handle the complex service Requests of the customers with the help of composition of two or more services as a new single service.

- Also the Reconfiguration of the system takes place at the run-time.

- Without interrupting the current running system we can easily insert a new composed service as a Feature Module.

- With the help of the case-based learning design pattern we can choose perfect decisions about the service plan to be accomplished to fulfill the Requested service.

## 4.8 Related Design Patterns

Strategy Design Pattern [8]: This pattern can be used to define a family of algorithms, encapsulate each one, and make them interchangeable. Strategy lets the algorithm vary independently from the clients that use it. In our proposed design pattern we use this pattern to choose or make a decision about the plan to be selected to fulfill the customer request or can be used as a decision-making pattern.

Adaptation Detector Design Pattern [15]: This design pattern can be used to interpret monitoring data and determine when an adaptation is required. With the help of observer design pattern it can monitor the system/application for any changes in the environment. Whenever any changes are detected then it will generate the Event as a Trigger which is then handled by Case-Based Reasoning pattern.

Architecture-Based Design Pattern [15]: This design pattern provides an architectural of selecting reconfiguration plans.

## 5. ROLES OF OUR DESIGN PATTERNS IN AUTONOMIC SYSTEM

- Case-Based Reasoning Design Pattern [15]: The Case-Based Reasoning Design Pattern in will apply the rule based decision making mechanism to determine a correct reconfiguration plan. This design pattern will separate the decision-making logic from the functional logic of the Application. In our proposed pattern we will use this pattern to provide the perfect suitable plan to implement the customer requested service.

- Reactor Design Pattern: Reactor Design Pattern in [6] handles service requests that are received concurrently from more than one client. In our proposed pattern we use this design pattern for efficiently handling the requested services. Each service in an application may consist of and is represented by a separate event-handler that dispatches the service-specific request. So this task of dispatching is done by initiation dispatcher, which itself manages the registered event handlers.

# 6. FEATURE BASED SERVICE INSERTION INTO SERVER REPOSITORY WITH ASSOCIATION RULE BASED LEARNING

The main concept in our proposed pattern is that with the help of already stored information in the Learning repository, we can use this information and then use the Association Rule based Data Mining method to perform data mining operation on this information. With this type of operation we can derive the new services from the already existing services. But the newly derived services are a composition of two or more services. So with the help of this association rule based data mining we can include the new services into the Service Repository as a Aspectual Feature Module [2]. With this we can provide composition of services to fulfill the complex service requests of the customers.

The view of our proposed design pattern for the unstructured peer-to-peer computing System can be seen in the form of a class diagram see Figure 2.

# 7. THE APRIORI ALGORITHM FOR ASSOCIATION RULE BASED LEARNING

Here we will provide the pseudocode for the frequent itemset generation part of the Apriori algorithm with reference from [20]. We have implemented the code for this algorithm and used in our application.

1: k=1.
2: $F_k$={i | $i \epsilon I \wedge \sigma(\{i\}) \geq N \times minsup$}.$\{Find\,all\,frequent\,1-itemsets\}$
$3: repeat$
$4: k = k + 1.$
$5: C_k$=apriori-gen($F_{k-1}$). //Generate candidate itemsets
6: for each transaction t $\epsilon T do$
$7: C_t$= subset($C_k$,t). //Identify all candidates that belong to t
8: for each candidate itemset c $\epsilon C_t$ do
$9: \sigma(c) = \sigma(c) + 1.//Increment\,support\,count$
$10: \ endfor$
$11: \ endfor$
$12: F_k$= {c | $c \epsilon C_k \wedge \sigma(c) \geq N \times mibsup$}.$//Extract\,the\,frequent\,k-itemsets$
$13: until F_k$= $\phi$
$14: Result = \cup F_k.$

# 8. PROFILING RESULTS

We are presenting the profiling results taken for ten runs without applying this pattern and after applying this pattern using the profiling facility available in the Netbeans IDE. The graph is plotted taking the time of execution in milliseconds on Y-axis and the run count on the X-axis.

The graph has shown good results while executing the code with patterns and is shown in Figure 3.This can confirm the efficiency of the proposed pattern.

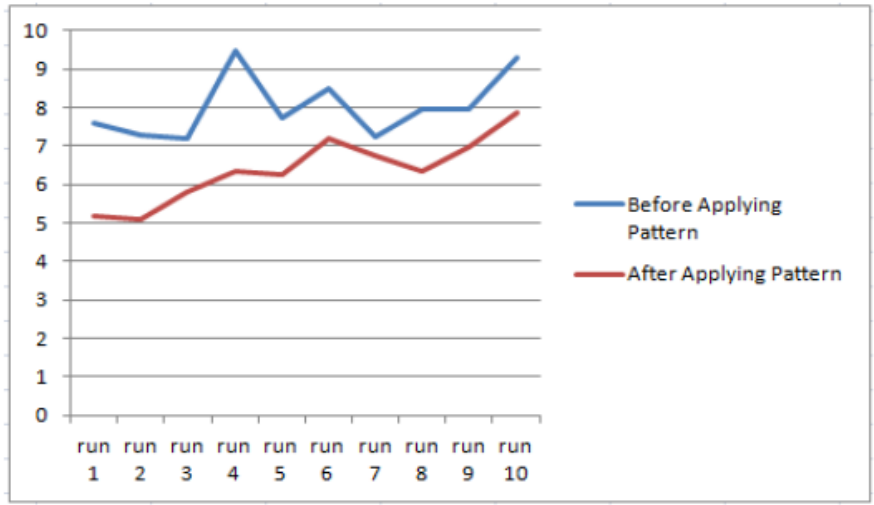

Figure 3: Profiling data after and before applying design pattern

# 9. SIMULATION RESULTS AFTER APPLYING SERVICE INJECTION DESIGN PATTERN

In our implementation we can evaluate the effectiveness of our implemented case-study with SOA and Socket Programming for web services invocation. In order to make our proposal clear we have successfully developed some critical parts of our system i.e., implementation of Case-Based Reasoning design pattern related code integrated with SOA with the help of Aspect oriented programming and feature oriented programming code and at the same time we have implemented the same application without design pattern (only some modules).

The simulation results for the code developed to prove the benefits of the proposed design pattern are collected with respect to:

- Used Heap memory
- Process CPU Time

# 10. DISCUSSION

From the Figure 7 and Figure 6 we can evaluate that the amount of Heap Memory used by applying code written without using design patterns is 13.8 M and where as for the amount of Heap Memory used with AOP and FOP based application Development along with SOA is 13.3 M. It is clear that the application developed using AOP along with SOA takes less heap memory when compared to implementation with respect to without design pattern.

From the Figures 5 and Figure 4 we can evaluate that the amount of CPU Time used by applying AOP and FOP based application Development along with SOA is 27% and where as for the amount of CPU Time used by applying Object Oriented implemented code without design pattern is 38%. It's clear that the application developed using AOP along with SOA takes less CPU Time when compared to implementation without using design pattern.

The view of our proposed design pattern for the Distributed SOA based computing System can be seen in the form of a class diagram see Figure 2.

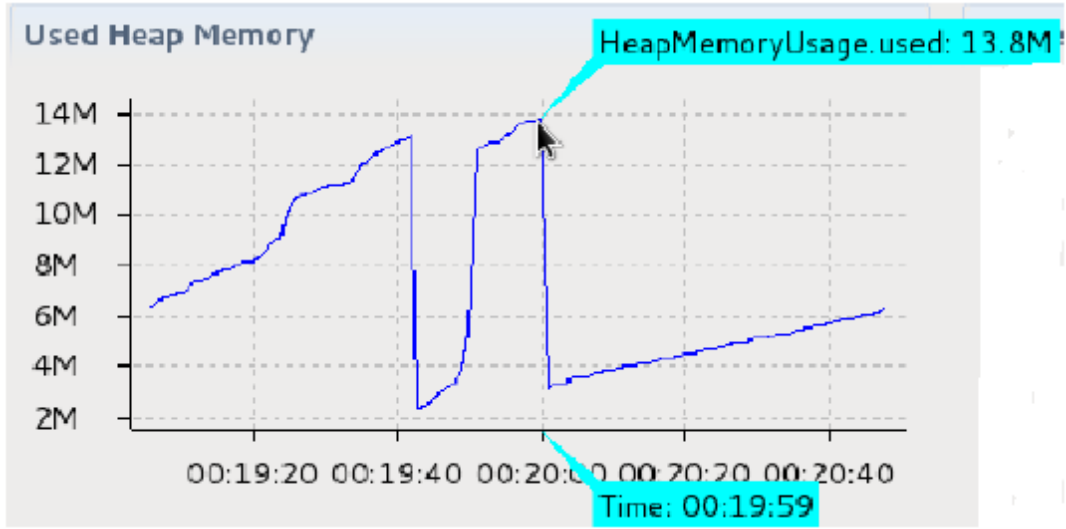

Figure 4: The CPU usage level is show above after applying our design pattern

# 11. CONCLUSION AND FUTURE WORK

In this paper we have proposed a pattern to facilitate Dynamic Service Composition. Here we have shown how the Association rule based Learning method of Data Mining can be used to determine the new services which are the composition of two or more services. Then we have also studied how the new services can be inserted into the service repository as a Aspectual Feature Module. Several future directions of work are possible. We are examining how these design patterns can be applied in the Software Product Lines (SPL). Also focusing upon the efficient use of the Data Mining methods for providing the flexible use of the design patterns for the most recurring problems.

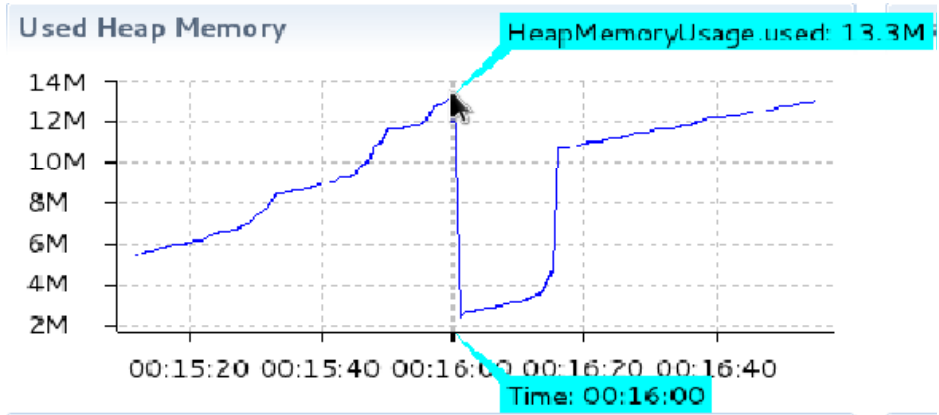

Figure 5: The CPU usage level is show above before applying our design pattern

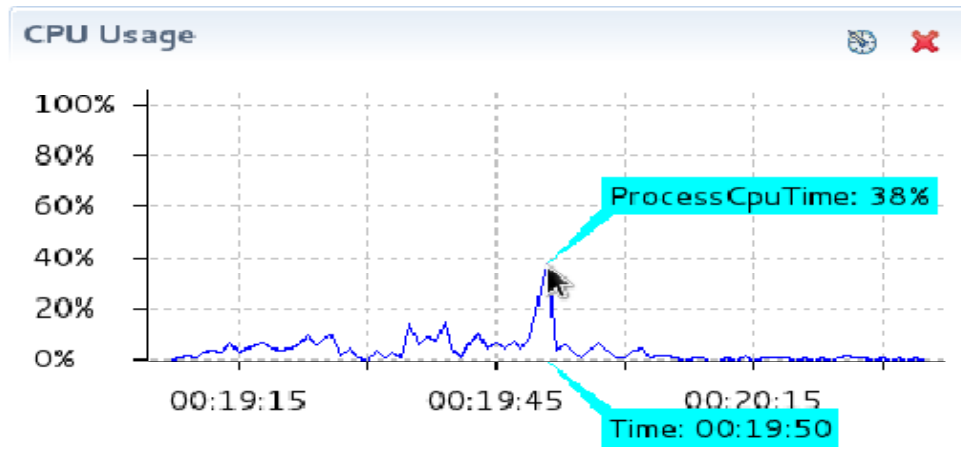

Figure 6: The Heap Memory usage shown for some critical module developed after applying our proposed design pattern.

## A. CASE STUDY FOR A SAMPLE APPLICATION DEVELOPED USING THE PROPOSED DESIGN PATTERN

In this section we would like to introduce a simple application which we have developed in order to authenticate our proposal of a dynamically adaptive design pattern. We have successfully developed some critical modules of our proposed design pattern and because of the huge code we are only limited to show small snippets of code. Here we will explain clearly what the each part of code does to provide the self-configuration capability to our developed system.

Initially we will present the Server side code and in the next section we will show some profiling and simulation results.

Here we have developed a simple electronic items purchase catalog and all the client transactions are stored in a database table named minethis. we have taken a sample of 6 items say laptops, external hard disks, pen drives, printers, modems, os dvd's as items that a client may purchase. So here after all the transaction data is stored in Mysql database we have applied the Appriori algorithm implementation on that stored data to get the frequent item sets of either 1-frequent item sets or 2-frequent item sets or 3-frequent item sets. Depending upon the support value set by default in a file with name config.txt which is used my the appriori algorithm to find the items that can be composed and given as new services that a user may ask for. So these new services are inserted as Feature modules into the current existing code with the help of fuji compliler used for Feature-Oriented Programming.

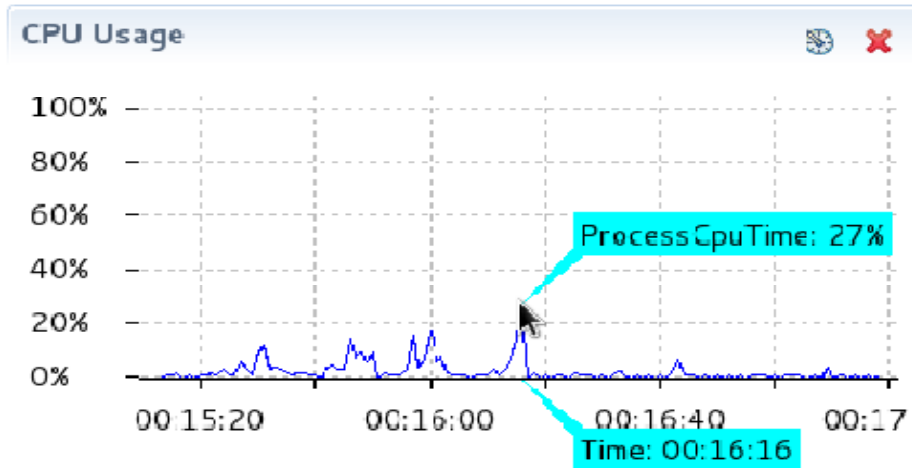

Figure 7: The Heap Memory usage shown for some critical module developed in without applying our proposed design pattern

Aspect Service Code is provided in AspectService3.aj

```
public abstract aspect Abstractservice3
{
Print p=new Print();
public abstract pointcut service3();
before():service3()
{
p.printbefore();
}
after():service3()
{
p.printafter();
}
}
```

Code for Inference Engine class

```
public class Infer_Engine {
public void notify(String arg1,String arg2,String arg3,String
arg4)
{
Fixed_Rules fr1=new Fixed_Rules();
```

```
System.out.println("Sending 4 items to Fixed
rules module for the purpose of Plan Selection");
fr1.rules(arg1,arg2,arg3,arg4);
}
public void notify(String arg1,String arg2)
{
Fixed_Rules fr2=new Fixed_Rules();
System.out.println("Sending 2 items to Fixed
rules module for the purpose of Plan Selection");
fr2.rules(arg1,arg2)
}
}
```

**Code for Test3.aj**

```
public aspect Test3 extends Abstractservice3
{
public pointcut service3() :
execution(* ThreeServicesImplement.Implementation1(. .) );
}
```

**Code for Decision Class**

```
public class Decision {
public void plan1(String arg1,String arg2,String arg3,String
arg4)
{
System.out.println("Here comes the Plan
to ur requested Service for item list of 4...");
System.out.println("Now the Reactor pattern
will handle Request Event of 4 items....");
Init_Dispacher id1=new Init_Dispacher(arg1,arg2,arg3,arg4);
}
public void plan2(String arg1,String arg2)
{
System.out.println("Here comes the Plan to
ur requested Service for item list of 2...");
System.out.println("Now the Reactor pattern
will handle Request Event of 2 items....");
Init_Dispacher id1=new Init_Dispacher(arg1,arg2);
}
}
```

The Transactions details that are given as input to appriori algorithm (these are the transactions collected after implementing our electronic catalog service) file name is transa.txt

```
1 1 1 1 0 0
1 1 0 0 0 0
1 0 1 1 1 0
0 1 1 1 0 1
1 1 1 0 0 1
```

The file congif.text have the information about number of items, number of transactions, support value (in percentage)

6
5
40

Code for Initiation Dispatcher class

```java
import java.awt.BorderLayout;
import java.awt.FlowLayout;
import java.awt.Font;
import java.awt.GridLayout;
import java.awt.event.ActionEvent;
import java.awt.event.ActionListener;
import java.sql.*;
import javax.swing.JButton;
import javax.swing.JFrame;
import javax.swing.JLabel;
import javax.swing.JPanel;
public class Init_Dispacher extends JFrame {
String item1=null;
String item2=null;
String item3=null;
String item4=null;
Connection con;
PreparedStatement pst;
ResultSet rs;
int total_cost;
public Init_Dispacher(String arg1,String arg2,String arg3,String
arg4)
{
item1=arg1;item2=arg2;item3=arg3;item4=arg4;
try{

//MAKE SURE YOU KEEP THE mysql_connector.jar file
in java/lib folder
//ALSO SET THE CLASSPATH
Class.forName("com.mysql.jdbc.Driver");
con=DriverManager.getConnection("jdbc:mysql://localhost
:3306/mydatabase","ataullah","ataullah");
pst=con.prepareStatement("select sum(price) from
Prices where item_name=? or item_name=? or
```

```
item_name=? or item_name=?");
pst.setString(1, item1); //this replaces the
1st "?" in the query for username
pst.setString(2, item2); //this replaces the
2st "?" in the query for password
pst.setString(3, item3); //this replaces the 1st
"?" in the query for username
pst.setString(4, item4); //this replaces the 2st
"?" in the query for password
//executes the prepared statement
rs=pst.executeQuery();
if(rs.next())
{
total_cost=rs.getInt(1);
}
}
catch (Exception e)
{
System.out.println(e);
}
setLayout(new GridLayout(3,1,5,10));
JLabel l1=new JLabel("Here comes your cost:"+total_cost);
JButton b1=new JButton("OK");
JButton b2=new JButton("Cancel");

JPanel p1=new JPanel();
p1.setLayout(new GridLayout(1,1,5,10));
p1.add(l1);

    JPanel p2=new JPanel();
    p2.setLayout(new FlowLayout());
    p2.add(b1);
    p2.add(b2);
    Font f=new Font("Algerian",Font.BOLD,20);
    JLabel tl=new JLabel("Plz... Conform whether
you want to buy the items");
    tl.setFont(f);
    JPanel p3=new JPanel();
    p3.add(tl);
    JPanel p5=new JPanel();
    p5.add(p1,BorderLayout.CENTER);
    p5.add(p2,BorderLayout.SOUTH);
    JPanel p4=new JPanel();
    p4.add(p3,BorderLayout.NORTH);
    p4.add(p5,BorderLayout.CENTER);
    //p4.add(p2,BorderLayout.SOUTH);

    add(p4);
    setVisible(true);
```

```
    setTitle("ATAULLA Quadri");
  setSize(150, 150);
    b1.addActionListener(new EventHandler(arg1,arg2,arg3,arg4));
    b2.addActionListener(new Terminate());
}
public Init_Dispacher(String arg1,String arg2)
{
item1=arg1;item2=arg2;
try{
Class.forName("com.mysql.jdbc.Driver");
con=DriverManager.getConnection("jdbc:mysql://
localhost:3306/mydatabase","ataullah","ataullah");
    pst=con.prepareStatement("select sum(price)
from Prices where item_name=? or item_name=?");
    pst.setString(1, item1);
pst.setString(2, item2);
//executes the prepared statement
rs=pst.executeQuery();
if(rs.next())
{
total_cost=rs.getInt(1);
}
  }
catch (Exception e)
{
System.out.println(e);
}
setLayout(new GridLayout(3,1,5,10));
JLabel l1=new JLabel("Here comes your total cost for
"+item1+"and"+item2+"is:"+total_cost);
JButton b1=new JButton("OK");
JButton b2=new JButton("Cancel");
JPanel p1=new JPanel();
p1.setLayout(new GridLayout(1,1,5,10));
p1.add(l1);
    JPanel p2=new JPanel();
    p2.setLayout(new FlowLayout());
    p2.add(b1);
    p2.add(b2);
    Font f=new Font("Algerian",Font.BOLD,20);
    JLabel tl=new JLabel("Plz... Conform
whether you want to buy the items");
    tl.setFont(f);
    JPanel p3=new JPanel();
    p3.add(tl);
    JPanel p5=new JPanel();
    p5.add(p1,BorderLayout.CENTER);
    p5.add(p2,BorderLayout.SOUTH);
    JPanel p4=new JPanel();
```

```
    p4.add(p3,BorderLayout.NORTH);
    p4.add(p5,BorderLayout.CENTER);
    //p4.add(p2,BorderLayout.SOUTH);
    add(p4);
    setVisible(true);
    setTitle("ATAULLA Quadri");
setSize(150, 150);
    b1.addActionListener(new EventHandler(arg1,arg2));
    b2.addActionListener(new Terminate());
}
}
```